\documentclass{article}
\usepackage{amsmath,amsfonts,xypic}
\usepackage{epsfig,psfig}
\usepackage{amssymb}
\usepackage{amscd}
\usepackage[all]{xy}

\newtheorem{theorem}{Theorem}[section]

\newtheorem{proposition}[theorem]{Proposition}
\newtheorem{definition}{Definition}[section]

\newcommand{\beqa}{\begin{eqnarray}}
\newcommand{\eeqa}{\end{eqnarray}}
\newcommand{\noi}{\noindent}
\newcommand{\e}{\varepsilon}

\newcommand{\om}{\Omega}
\newcommand{\tom}{\tilde \Omega}
\newcommand{\qed}{{\hfill $\Box$}}

\begin{document}

\title{
{\bf Dimensional regularization and renormalization of non-commutative QFT}  }

\author{ 
{\sf   R. Gurau}\thanks{e-mail:
razvan.gurau@th.u-psud.fr}$\,\,$ $^{}$
and
{\sf A. Tanas\u a}\thanks{e-mail:
adrian.tanasa@ens-lyon.org}$\,\,$ $^{}$
\\
{\small ${}^{a}${\it Laboratoire de Physique Th\'eorique, CNRS UMR 8627,}} \\
{\small {\it b\^at. 210, 
Universit\'e Paris XI, 91405 Orsay Cedex, France.}}  \\ 
}
%\date{\today}
\maketitle

\medskip

\begin{abstract}
\noindent
Using the recently introduced parametric representation of non-commutative quantum field theory, we implement here the dimensional regularization and renormalization of the vulcanized $\Phi^{\star 4}_4$ model on the Moyal space.
\end{abstract}


\section{Introduction and motivation}
\setcounter{equation}{0}

Non-commutative geometry (see \cite{connes}) is one of the
most appealing frameworks for the quantification of gravitation.
Quantum field theory (QFT) on these type of spaces, called non-commutative quantum
field theory (NCQFT) - for a general review see
\cite{nc1,nc2}- is now
one of the most appealing candidates for new physics beyond the Standard
Model. Also, NCQFT arises as the effective limit of some string
theoretical models \cite{cordes1, cordes2}.

Moreover, NCQFT is well suited to the description of the physics in background fields and with non-local interactions, like for example the
fractional quantum Hall effect \cite{hall1, hall2, hall3}.

However, naive NCQFT suffers from a new type of non renormalizable divergences, known as the ultraviolet (UV)/ infrared (IR) mixing. The simplest example of this kind of divergences is given by the nonplanar tadpole: it is UV convergent, but inserting it an arbitrary number of times in a loop gives rise to IR divergences.

Interest in NCQFT has been recently revived with the introduction
of the Grosse-Wulkenhaar scalar $\Phi^{\star 4}_4$ model, in which the UV/IR mixing is cured: the model is
renormalizable at all orders in perturbation theory \cite{GW1,GW2}. The idea of Grosse-Wulkenhaar was to modify the kinetic part of the action in order to satisfy the Langmann-Szabo duality \cite{ls} (which relates the infrared and
ultraviolet regions). We refer to this modified theory as the vulcanized 
$\Phi^{\star 4}_4$ model.

A general proof, using position space and multiscale analysis, has then been given in \cite{4men}, and the parametric representation of this model was computed in \cite{param1}. Furthermore, it was recently proved
that that the vulcanized $\Phi^{\star 4}_4$ is better behaved than
the commutative $\phi^4_4$ model: it does {\bf not} have a Landau ghost \cite{landau1, landau2, landau3}.

In commutative QFT dimensional renormalization is the only scheme which respects the symmetries of gauge theories see \cite{reg, ren, nemtii}. It also is the appropriate setup for the Connes-Kreimer Hopf algebra approach to renormalization (see 
\cite{ck1, ck2,kreimer} for the case of commutative QFT).
 
 A second class of renormalizable NCQFT exists. These models, called {\it covarinat}, are characterized
 by a propagator which decays in position space as $x-y$ tends to infinity (like the Grosse-Wulkenhaar propagator
of eq. (\ref{propa})) but it oscillates when $x+y$ goes to infinity, rather than decaying. In
this class of NCQFT models enters the
non-commutative Gross-Neveu model and the Langmann-Szabo-Zarembo model
\cite{LSZ}. The non-commutative orientable Gross-Neveu model was proven to be
renormalizable at any order in perturbation theory \cite{fab}. The
parametric representation was extended to this class of models \cite{param2}.
For a general review of recent developments in the field of renormalizable NCQFT see \cite{sefu}.

The parametric representation introduced in \cite{param1} is the starting point for the dimensional regularization and renormalization performed in this paper. Our proof follows that of the commutative $\Phi^{\star 4}_4$ model, as presented in \cite{reg, ren}.

This paper is organized as follows. Section \ref{sec:NCPHI44} is a summary of the parametric representation
of the vulcanized $\Phi^{\star 4}_4$ model. The non-commutative equivalent $HU_G$ and $HV_G$ of the Symanzik polynomials $U_G$
and $V_G$ are recalled.
In section \ref{sec:furtherLeading} we prove the existence in the polynomial $HU_G$ of some further leading terms in the ultraviolet (UV) regime.  This is an improvement of the results of \cite{param1}, needed to correctly identify the meromorphic structure of the Feynman amplitudes. 
In section \ref{sec:factorisation} we prove the factorization properties of the Feynman amplitudes. These properties are needed in order to prove that the pole extraction is equivalent to adding counterterms of the form of the initial lagrangean. This factorization is essential for the definition of a coproduct $\Delta$ necessary for the implementation of a Hopf algebra structure in NCQFT \cite{progres}. 
Section \ref{sec:NCdimreg} uses the results of the previous sections to perform the dimensional regularization, prove the counterterm structure for NCQFT and complete the dimensional renormalization program. 
Section \ref{sec:conclusion} is devoted to some conclusion and perspectives. 

\section{The non-commutative model}
\label{sec:NCPHI44}
\setcounter{equation}{0}

In this section we give a brief overview of the Grosse-Wulkenhaar $\Phi^4$ model. Our notations and conventions as well as some notions of diagrammatics and the results of the parametric representation follow \cite{param1}.

To define the Moyal space of dimension $D$, we introduce the deformed Moyal product $\star$ on ${\mathbb R}^D$ so that
\beqa
\label{2D}
[x^\mu, x^\nu]=i \Theta^{\mu \nu},
\eeqa
\noi
where the the matrix $\Theta$ is
\begin{eqnarray}
\label{theta}
  \Theta= 
  \begin{pmatrix}
    \begin{matrix} 0 &\theta \\ 
      \hspace{-.5em} -\theta & 0
    \end{matrix}    &&     0
    \\ 
    &\ddots&\\
    0&&
    \begin{matrix}0&\theta\\
      \hspace{-.5em}-\theta & 0
    \end{matrix}
  \end{pmatrix}.
\end{eqnarray}
\noi
The associative Moyal product of two functions 
$f$ and $g$ on the Moyal space writes
\beqa
\label{moyal-product} 
 (f\star g)(x)&=&\int \frac{d^{D}k}{(2\pi)^{D}}d^{D}y\, f(x+{\textstyle\frac 12}\Theta\cdot
  k)g(x+y)e^{\imath k\cdot y}\nonumber\\
  &=&\frac{1}{\pi^{D}|\det\Theta|}\int d^{D}yd^{D}z\,f(x+y)
  g(x+z)e^{-2\imath y\Theta^{-1}z}\; .
\eeqa
The Euclidian action introduced in \cite{GW2} is 
\beqa
\label{lag-init}
S=\int d^4 x \left(\frac{1}{2} \partial_\mu \phi
\star \partial^\mu \phi +\frac{\Omega^2}{2} (\tilde{x}_\mu \phi )\star
(\tilde{x}^\mu \phi ) 
+ \frac{1}{2} m^2 \,\phi \star \phi
+  \phi \star \phi \star \phi \star\phi \right)\, ,
\eeqa
\noi
where 
\beqa
\label{tildex}
\tilde{x}_\mu = 2 (\Theta^{-1})_{\mu \nu} x^\nu \, .
\eeqa
\noi
The propagator of this model is the inverse of the operator
\beqa
\label{inverse-propa}
-\Delta+\Omega^{2}\tilde x^{2}.
\eeqa
\noi
The results we establish here hold for orientable models (in the
sense of subsection \ref{grafuri}). This corresponds to a Grosse-Wulkenhaar
model of a complex scalar field
\beqa
\label{lag}
S=\int d^4 x \left(\frac{1}{2} \partial_\mu \bar \phi
\star \partial^\mu \phi +\frac{\Omega^2}{2} (\tilde{x}_\mu \bar \phi )\star
(\tilde{x}^\mu \phi ) 
+  \bar \phi \star \phi \star \bar \phi \star \phi \right) \, .
\eeqa
\noi
Introducing $\tom=2\Omega/\theta$, the kernel of the propagator is (Lemma $3.1$ of \cite{propagatori}) 
\beqa
\label{propa}
C(x,y)=\int_0^\infty \frac{\tom d\alpha}{[2\pi\sinh(\alpha)]^{D/2}}
e^{-\frac{\tom}{4}\coth(\frac{\alpha}{2})(x-y)^2-
\frac{\tom}{4}\tanh(\frac{\alpha}{2})(x+y)^2}\; .
\eeqa

Using eq. (\ref{moyal-product}) the interaction term in eq. (\ref{lag})
leads to the following vertex contribution in position space (see \cite{4men}
\beqa
\label{v1}
\delta (x_1 - x_2 + x_3 - x_4)e^{2i\sum_{1\le
    i <j\le 4}(-1)^{i+j+1}x_i\Theta^{-1}x_j} \, .
\eeqa
\noi
with $x_1,\ldots, x_4$ the $4-$vectors of the positions of the $4$
fields incident to the vertex.

To any such vertex $V$ one associates 
a hypermomentum $p_V$ using the relation
\beqa
\label{pbar1}
\delta(x_1 -x_2+x_3-x_4 ) 
= \int  \frac{d p_V}{(2 \pi)^4}
e^{p_V \sigma (x_1-x_2+x_3-x_4)} \, .
\eeqa
\noi

\subsection{Some diagrammatics for NCQFT; orientability}
\label{grafuri}

In this subsection we introduce some useful conventions and  definitions, some of them used in \cite{4men} and  \cite{propagatori} but also some new ones.

Let a graph $G$ with $n(G)$ vertices, $L(G)$ internal lines and $F(G)$
faces. The Euler characteristic of the graph is
\beqa
\label{genus}
2-2g(G)=n(G)-L(G)+F(G),
\eeqa
\noi
where $g(G)\in{\mathbb N}$ is the {\it genus} of the graph.
Graphs divide in two categories, {\it planar graph} with $g(G)=0$, and 
{\it non-planar graphs} with $g(G)>0$.
Let also $B(G)$ denote the number of faces broken by external lines and
$N(G)$ be number of external points of the graph.

The ``orientable'' form eq .(\ref{v1}) of the
vertex contribution of our model allows us to associate ``+'' sign to a corner $\bar \phi$ and a  ``-'' sign to a corner $\phi$ of the vertex.
These signs alternate when turning around a vertex.
As the propagator allways relates a $\bar \phi$ to a $\phi$, the action in eq. (\ref{lag}) has orientable lines, that is any internal line joins a ``-'' corner to a ``+'' corner \footnote{The orientability of our theory allows us to simplify the proofs. It should however be possible, although tedious, to follow the same procedure for the non orientable model.}.

Consider a spanning tree $ {\cal T} $ in $G$. In has $n-1$ lines and 
the remaining $L-(n-1)$ lines form the set $\cal L$ of loop lines. 
Amongst the vertices $V$ one chooses a special one $\bar V_{G}$, the {\it
  root} of the tree. 
One associates to any vertex $V$ the unique tree line which hooks to $V$ and goes towards the root. 

We introduce now some topological operations on the graph which allow one to reexpress the oscillating factors coming from the vertices of the graph $G$.

\begin{figure}
  \centerline{\epsfig{figure=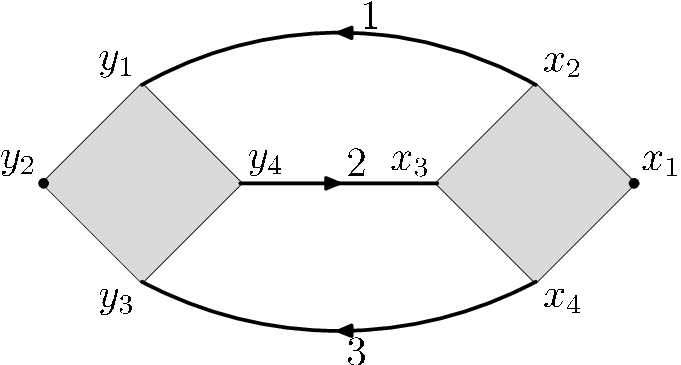,width=6cm} \hfil \epsfig{figure=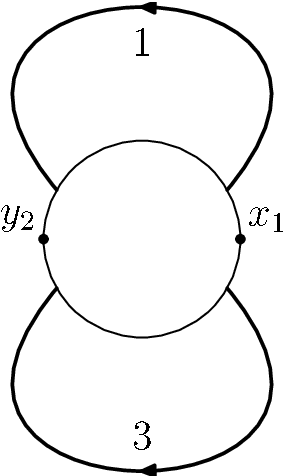,width=2cm}}
\caption{The first Filk move: the line $2$ is reduced and the $2$ vertices merge}\label{firstfilk}
\end{figure}

Let a tree line in the graph $\ell=(i,j)$ and its endpoints $i$ and $j$. Suppose it connects to the root vertex $\bar V_G$ at $i$ and to another vertex $V$ at $j$.
In Fig.\ref{firstfilk}, $\ell_2$ is the tree line, $y_4$ is $i$ and $x_1$ is $j$.
{\bf The first Filk move}, inspired by \cite{filk}, consists in removing such a 
line from the graph and gluing the two vertices together respecting the ordering. Thus the point $i$ on the root vertex is replaced by the neighbors of $j$ on $V$. 
This is represented in Fig. \ref{firstfilk} where the new root vertex is
$y_2,y_3,x_4,x_1,x_2,y_1$.

Note that the number of faces or the genus of the graph do not change 
under this operation.

A technical point to be noted here is that one must chose the field $j$ on the vertex $V$ to be either the first (if the line $\ell$ enters $V$) or the last 
(if the line $\ell$ exits $V$) in the ordering of $V$. Of course this is allways possible by the use of the $\delta$ functions in the vertex contribution.

\begin{figure}[ht]
\centerline{\epsfig{figure=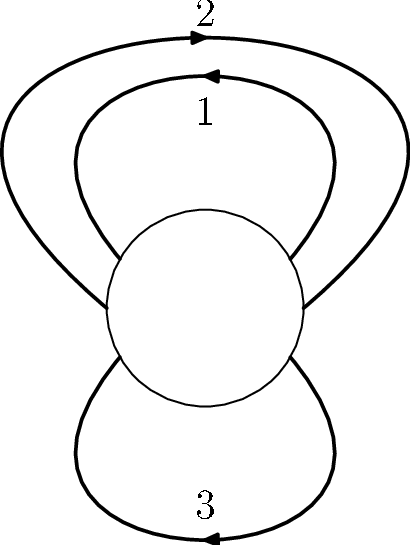,width=3cm}\hfil
  \epsfig{figure=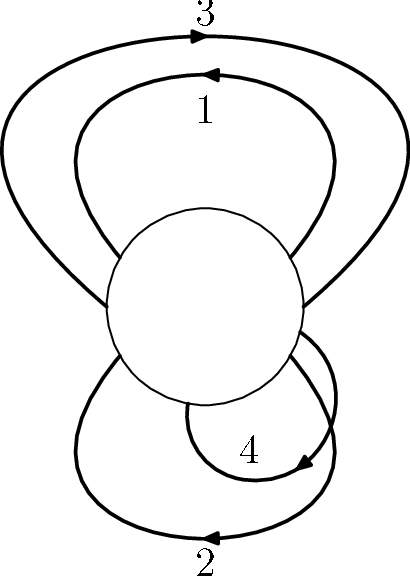,width=3cm}}
\caption{Two rosettes obtained by contracting a tree {\it via} the first Filk}\label{rozeta}
\end{figure}

Iterating this operation for the $n-1$ tree lines, one obtains a single final
vertex with all the loop lines hooked to it - a {\it rosette} (see Fig. \ref{rozeta}).

The rosette contains all the topological information of the graph. If no two lines cross (on the left in Fig. \ref{rozeta}) the graph is planar. If on the contrary we have at least a crossing (on the right in Fig. \ref{rozeta}) the graph is non planar (for details see \cite{param1}).

For a nonplanar graph we define a {\it nice crossing} in a rosette as a pair of lines such that the end point of the first is the successor in the rosette of the
starting point of the other. A {\it genus line} of a graph is a loop line which is part of a nice crossing on the rosette (lines $2$ and $4$ on the right of Fig. \ref{rozeta}).

In the sequel we are interested in performing this operation in a way adapted to the scales introduced by the Hepp sectors: we  perform the first Filk move only for a subgraph $S$ (we iterate it only for a tree in $S$).
Thus, the subgraph $S$ will be shrunk to its
corresponding rosette inside the graph $G$. If $S$ is not primitively divergent we have a convergent sum over its associated Hepp parameter.

We will prove later that $S$ is primitively divergent if and only if $g(S)=0$, $B(S)=1$, $N(S)=2,4$.
For primitively divergent subgraphs the first Filk move above shrinks $S$ to a Moyal vertex inside the graph
$G$. 

\begin{figure}[ht]
\centerline{\epsfig{figure=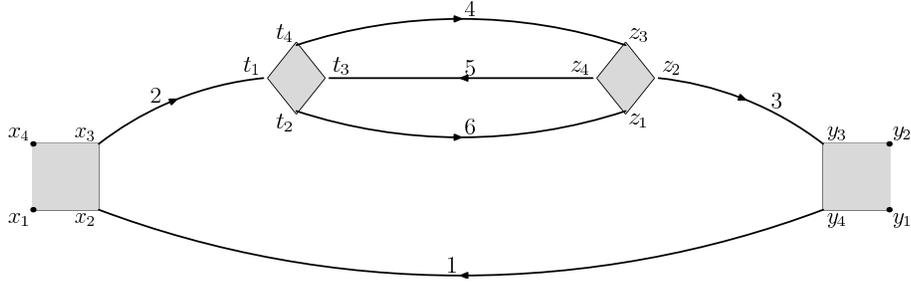,width=12cm}}
\caption{A graph containing a primitive divergent subgraph given by the lines  $\ell_4,\ell_5$ and $\ell_6$}\label{hyper}
\end{figure}

\begin{figure}[ht]
\centerline{\epsfig{figure=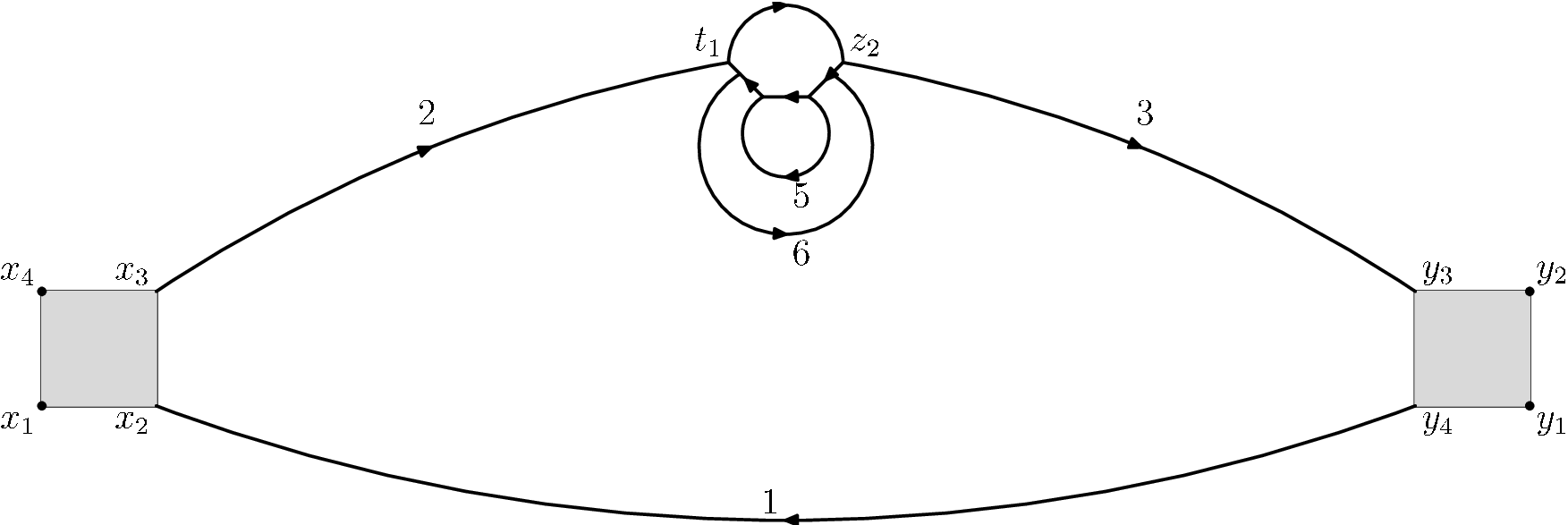,width=12cm}}
\caption{The graph $G$ with $S$ reduced to a rosette}
\label{fig:mese}
\end{figure}

\begin{figure}[ht]
\centerline{\epsfig{figure=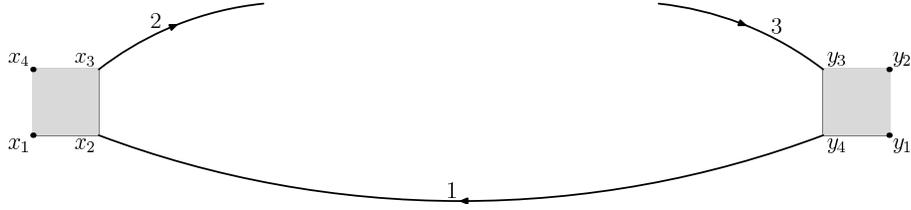,width=12cm}}
\caption{The graph $G-S$ obtained by erasing the lines
 and vertices of the primitive divergent subgraph $S$}
\label{G-S}
\end{figure}

\begin{figure}[ht]
\centerline{\epsfig{figure=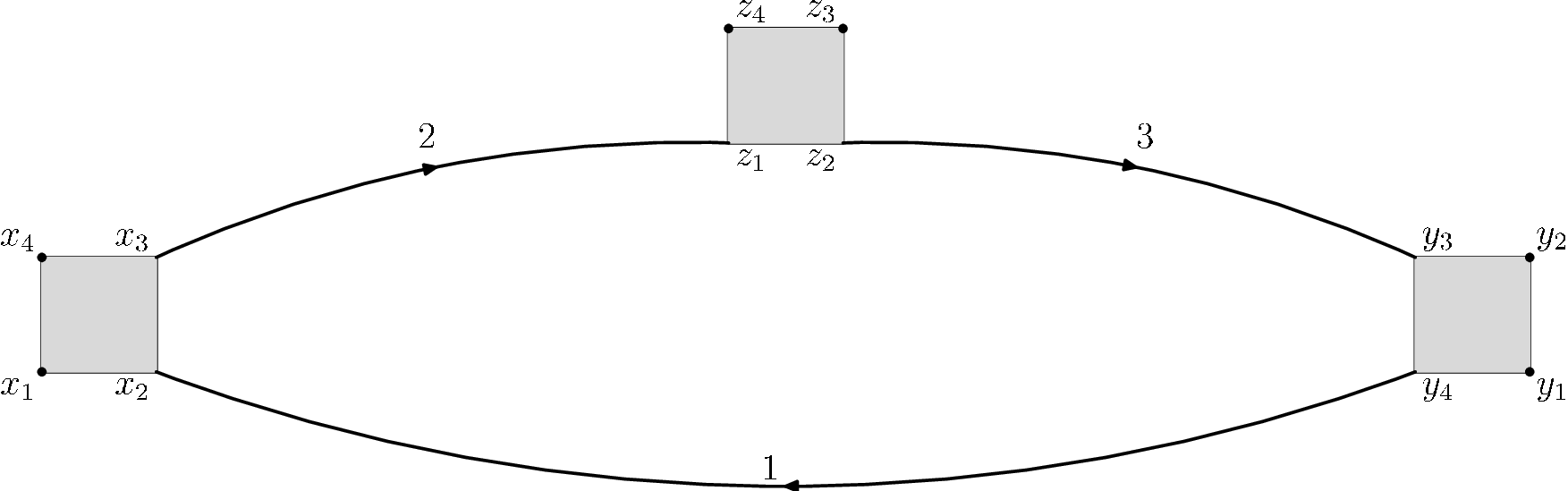,width=12cm}}
\caption{The graph $G/S$ obtained by shrinking to a Moyal vertex  the sunshine
  primitive divergent subgraph $S$}
\label{bula-deformata}
\end{figure}

For example, consider the graph $G$ of Fig. \ref{hyper} and its divergent
sunshine subgraph $S$ given by the set of lines $\ell_4,\ell_5$ and $\ell_6$. 
Under the first Filk move for the subgraph $S$, $G$  will have a rosette vertex insertion like in Fig. \ref{fig:mese},
Denote $G-S$ graph $G$ with its subgraph $S$ erased (see Fig.  \ref{G-S}). It becomes the graph $G/S$ with a Moyal vertex like in Fig. \ref{bula-deformata}.

In the commutative case, this operation corresponds to the shrinking of $S$ to a point: it represents the ''Moyality" (instead of locality) of the theory.

\subsection{Parametric representation for NCQFT}
\label{para-nc}

In this subsection we recall the definitions and results obtained in
\cite{param1} for the parametric
representations of the model defined by eq. (\ref{lag}).
First let us recall that, when considering the parametric
representation for commutative QFT, one has translation invariance in position
space. As a consequence of this invariance, the first polynomial vanishes when
integrating over all internal positions. Therefore, one has to integrate over
all internal positions (which correspond to vertices) save one, which is thus
marked. However, the polynomial is a still a canonical object, {\it
  i. e.} it {\it does not depend} of the choice of this
particular vertex.

As stated in \cite{param1}, in the non-commutative case translation
invariance is lost (because of non-locality). Therefore, one can integrate
over all internal positions and hypermomenta. However, in order to be able to recover the commutative limit, we
also mark a particular vertex $\bar V$; we do not integrate on its associate hypermomenta $p_{\bar V}$. This particular vertex is the root vertex. Because there is no translation invariance, the polynomial does depend on on the choice of the root; however the leading ultraviolet terms do not.

We define the $(L\times 4)$-dimensional incidence matrix $\e^V$ for each of the vertices $V$. Since the graph is orientable (in the sense defined in subsection
\ref{grafuri} above) we can choose
\beqa
\label{r1}
\e_{\ell i}^V= (-1)^{i+1}, \mbox { if the line $\ell$ hooks to the vertex $V$
  at corner $i$.}
\eeqa
\noi
Let also
\beqa
\label{r2}
\eta^V_{\ell i}=\vert \e^V_{\ell i}\vert, \mbox { } V=1,\ldots, n,\, 
\ell=1,\ldots, L \mbox{ and } i=1,\ldots, 4. 
\eeqa
From eq . (\ref{r1}) and (\ref{r2}) one has
\beqa
\label{r3}
\eta^V_{\ell i} = (-1)^{i+1}\e_{\ell i}.
\eeqa
We introduce withe the ''short" $u$ and ''long" $v$ variables by 
\beqa
v_\ell&=&\frac{1}{\sqrt{2}} \sum_V \sum_i \eta^V_{\ell i} x^V_i,\nonumber\\
u_\ell&=&\frac{1}{\sqrt{2}} \sum_V \sum_i \e^V_{\ell i} x^V_i.
\eeqa
Conversely, one has
\beqa
 x^V_i= \frac{1}{\sqrt{2}}\left(\eta^V_{\ell i}v_\ell+\e^V_{\ell i}u_\ell \right).
\eeqa

From the propagator \ref{propa} and vertices
contributions \ref{v1} one is able to write the amplitude ${\cal A}_{G,{\bar
    V}}$ of the graph $G$ (with the marked root $\bar V$) in terms of
the non-commutative polynomials $HU_{G, \bar{V}}$ and $HV_{G, \bar{V}}$ as (see
\cite{param1} for details)
\beqa
\label{HUGV}
{\cal A}_{G,{\bar V}}  (x_e,\;  p_{\bar V}) = \left(\frac{\tom}{2^{\frac D2
      -1}}\right)^L  \int_{0}^{\infty} \prod_{\ell=1}^L  [ d t_\ell
(1-t_\ell^2)^{\frac D2 -1} ]
\frac
{e^{-  \frac {HV_{G, \bar{V}} ( t_\ell , x_e , p_{\bar v})}
{HU_{G, \bar{V}} ( t )}}}
{HU_{G, \bar{V}} ( t )^{\frac D2}},
\eeqa
with $x_e$ the external positions of the graph and 
\beqa
\label{t}
t_\ell = {\rm tanh} \frac{\alpha_\ell}{2}, \ \ell=1,\ldots, L.
\eeqa
where  $\alpha_\ell$ are the parameters associated by eq. (\ref{propa}) to
the propagators of the graph. 
In \cite{param1} it was proved that  $HU$ and $HV$ are polynomials in the set of variables $t$. The first polynomial is given by (see again \cite{param1})
\beqa
\label{huqv12}
HU_{G, \bar{V}}=({\rm det} Q)^{\frac 1D} \prod_{\ell=1}^L t_\ell \, ,
\eeqa
where
\beqa
\label{Q}
Q= A\otimes 1_D - B \otimes \sigma \, ,
\eeqa
with $A$ a diagonal matrix and $B$ an antisymmetric matrix. The matrix $A$ writes 
\beqa 
\label{defmatrixa}
 A=\begin{pmatrix} S & 0 & 0\\ 0  & T & 0 \\ 0&0&0\\
\end{pmatrix} \, ,
\eeqa 
where $S$ and resp. $T$ are the two diagonal $L$ by $L$ matrices
with diagonal elements $c_\ell = \coth(\frac{\alpha_\ell}{2}) = 1/t_\ell$, 
and resp. $t_\ell$.
The last $(n-1)$ lines and columns are have $0$ entries.

The antisymmetric part $B$ is
\beqa
\label{b}
B= \begin{pmatrix}{s} E & C \\
-C^t & 0 \\
\end{pmatrix}\, , 
\eeqa
with 
\beqa
s=\frac{2}{\theta\tom}=\frac{1}{\om}\, ,
\eeqa
and 
\beqa
\label{c}
C_{\ell V}=\begin{pmatrix}
\sum_{i=1}^4(-1)^{i+1}\epsilon^V_{\ell i} \\
\sum_{i=1}^4(-1)^{i+1}\eta^V_{\ell i} \\
\end{pmatrix}\ ,
\eeqa
\beqa
\label{e}
E=\begin{pmatrix}E^{uu} & E^{uv} \\ E^{vu} & E^{vv} \\
\end{pmatrix}.
\eeqa
The blocks of the matrix $E$ are
\beqa
\label{ee}
E^{vv}_{\ell,\ell'}&=&\sum_V
\sum_{i,j=1}^4  (-1)^{i+j+1} \omega(i,j)\eta_{\ell i}^V\eta_{\ell' j}^V,
\nonumber\\
E^{uu}_{\ell,\ell'}&=&\sum_V
\sum_{i,j=1}^4  (-1)^{i+j+1} \omega(i,j)\epsilon_{\ell i}^V\epsilon_{\ell' j}^V,
\nonumber\\
E^{uv}_{\ell,\ell'}&=&\sum_V
\sum_{i,j=1}^4  (-1)^{i+j+1} \omega(i,j)\epsilon_{\ell i}^V\eta_{\ell'
  j}^V. 
\eeqa
The symbol $\omega(i,j)$ takes the values $\omega(i,j)=1$ if $i<j$, 
$\omega(i,j)=-1$ is $j<i$ and $\omega(i,j)=0$ if $i=j$.
In eq. (\ref{c}) of the matrix $C$ we have rescaled by $s$ the hypermomenta $p_V$. 
For further reference we introduce the integer entries matrix:
\beqa
\label{b'}
B'= \begin{pmatrix} E & C \\
-C^t & 0 \\
\end{pmatrix}\ .
\eeqa

In \cite{param1} it was proven that
\beqa
\label{qm}
{\rm det} Q= ({\rm det} M)^D \, ,
\eeqa
where
\beqa
\label{M}
M= A+B \, .
\eeqa
Thus eq. (\ref{huqv12}) becomes:
\beqa
\label{hugvq2}
HU_{G, \bar{V}}={\rm det} M \prod_{\ell=1}^L t_\ell \, .
\eeqa

Let $I$ and resp. $J$ be two subsets of $\{1,\ldots,L\}$, of cardinal $\vert I \vert$ and $\vert J \vert$. Let
\beqa
\label{kij}
k_{I,J} = \vert I\vert+\vert J\vert - L - F +1 \, ,
\eeqa
and $n_{I J}=\mathrm{Pf}(B'_{\hat{I}\hat{J}})$, the Pffafian of the  matrix
$B'$ with deleted lines and columns $I$ among the first $L$ indices 
(corresponding to short variables $u$) and $J$ among the next $L$ 
indices (corresponding to long variables $v$).

The specific form \ref{defmatrixa}
allows one to write the polynomial $HU$ as a sum of positive terms:
\beqa
\label{suma}
HU_{G,{\bar V}} (t) &=&  \sum_{I,J}  s^{2g-k_{I,J}} \ n_{I,J}^2
\prod_{\ell \not\in I} t_\ell \prod_{\ell' \in J} t_{\ell'}\ .
\eeqa

In \cite{param1}, non-zero {\it leading terms} ({\it i. e.} terms with
the smallest global degree in the $t$ variables) were identified. 
These terms are dominant in the UV regime. 
Some of them correspond to subsets $I=\{1,\ldots, L\}$ and $J$  {\it admissible}, that is 
\begin{itemize}
\item $J$ contains a tree $\tilde {\cal T}$ in the dual graph,
\item the complement of $J$ contains a tree $\cal T$ in the direct graph.
\end{itemize}
Associated to such $I$ and $J$ one has $n_{I,J}^2=2^{2g}$.

However, the list of these leading terms, as already remarked in
\cite{param1}, is not exhaustive. In the next section we complete this list
with further terms, necessary for the sequel \footnote{For the purposes of \cite{param1}, the existence of some non-zero leading
terms was sufficient.}.

\begin{figure}[ht]
\centerline{\epsfig{figure=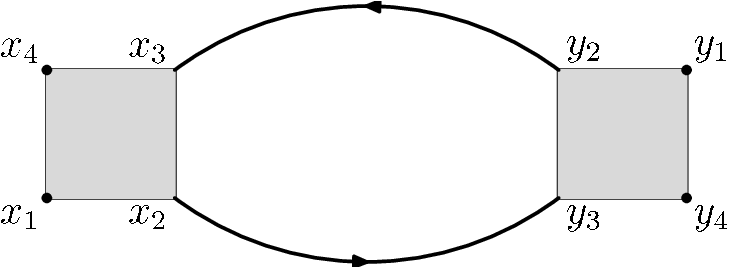,width=6cm}}
\caption{The bubble graph}\label{bubble}
\end{figure}

\begin{figure}[ht]
\centerline{\epsfig{figure=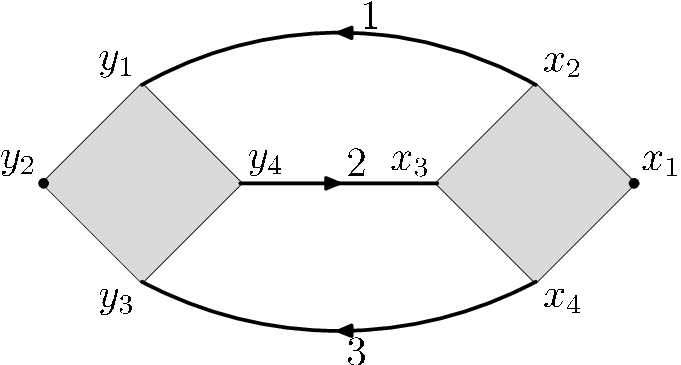,width=6cm}}
\caption{The sunshine graph}\label{sunshine}
\end{figure}

We end this section with the explicit example of the bubble and the sunshine graph (Fig. \ref{bubble} and Fig. \ref{sunshine}).

For the bubble graph one has
\beqa
\label{pol-bula}
HU_{G,\bar V}&=&(1+4s^2)(t_1+t_2+t_1^2t_2+t_1t_2^2).
\eeqa

For the sunshine graph one has
\beqa
\label{pol-sunshine}
HU_{G,\bar V}&=&\Big{[} t_1t_2+t_1t_3+t_2t_3+t_1^2t_2t_3+t_1t_2^2t_3+t_1t_2t_3^2\Big{]}
(1+4s^2)^2\nonumber\\
&&+16s^2(t_2^2+t_1^2t_3^2).
\eeqa

For further reference we also give the polynomial of the graph of Fig. \ref{bula-deformata}
\beqa
\label{bula-traficata}
HU_{G,\bar V}=(1+4s^2) (t_1 + t_2 + t_3 + t_1 t_2 t_3) (1+ t_2 t_3 + t_1 (t_2+ t_3)).
\eeqa

The polynomial $HV$ is more involved. One has
\beqa
\label{eq:HVGV}
\frac{HV_{G,\bar V}}{HU_{G,\bar V}}=\frac{\tom}{2}
\begin{pmatrix}x_e & p_{\bar V} \end{pmatrix}
PQ^{-1}P^t
\begin{pmatrix}x_e \\ p_{\bar V} \end{pmatrix},
\eeqa 
where $P$ is some matrix coupling the external positions $x_e$ and the root
hypermomenta $p_{\bar V}$ with the short $u$ and long $v$ variables and the
rest of the hypermomenta $p_V$ ($V\ne \bar V$). Explicit expressions can be found in \cite{param1}.

\section{Further leading terms in the first polynomial HU}
\label{sec:furtherLeading}
\setcounter{equation}{0}

To procede with the dimensional regularization one needs first to correctly isolate the divergent subgraphs in different Hepp sectors. Let a subgraph $S$ in the graph G. If $S$ is nonplanar the leading terms in $HU_G$ suffice to prove that $S$ is convergent in every Hepp sector as will be explained in section \ref{sec:NCdimreg}. 

This is not the case if $S$ is planar with more that one broken face. 
Let $\tilde \ell$ be the line of $G$ which breaks an internal face of $S$.

If  $\tilde \ell$ is a genus line in $G$ (see subsection \ref{grafuri}), one could still use only the leading terms in $HU_G$ to prove that $S$ is convergent.

But one still needs to prove that $S$ is convergent if the line $\tilde \ell$ is {\bf not} a genus line in $G$. This is for instance the case of the sunshine graph: in the Hepp sector $t_1< t_3 <t_2$ one must prove that the subgraph formed by the lines $l_1$ and $l_3$ is convergent. 
This is true due to the term $16 s^2 t_2^2$ in  eq. (\ref{pol-sunshine}).
Note that the variable $t_2$ is associated to the
line which breaks the internal face of the subgraph $\ell_1$, $\ell_3$.
One needs to prove that for arbitrary $G$ and $S$ we have such terms.

If we reduce $S$ to a rosette there exists a loop line $\ell_2\in S$ which 
 either  crosses $\tilde \ell$ or encompasses it. This line separates the two 
broken faces of $S$.

\begin{definition}
Let $J_0$ a subset of the internal lines of the graph $G$. $J_0$ is called
{\it pseudo-admissible} if: 
\begin{itemize}
\item its complement is the union of tree $\cal T$ in $G$ and $\ell_2$,
\item neither $\tilde \ell$ nor $\ell_2$ belong to $\cal T$,
\end{itemize}
\end{definition}

Let $I_0=\{\ell_1 \hdots \ell_L\} - \tilde \ell \equiv I- \tilde
\ell$. This implies $|I|=L(G)-1$ and $|J|=F(G)-2+2g(G)$. For the sunshine graph (see Fig. \ref{sunshine}) $I=\{ \ell_1, \ell_3 \}$ and $J=\{ \ell_2 \}$. 
One has the theorem

\begin{theorem}
\label{completare}
In the sum \ref{suma} the term associated to $I_0$ and $J_0$ above is
\beqa
n_{I_0, J_0}^2&=&4, \mbox{ if $\tilde \ell$ is a genus line, in $G$}\nonumber\\
&=& 16, \mbox{ if $\tilde \ell$ is not a genus line in $G$.}\nonumber
\eeqa
\end{theorem}
{\bf Proof:} 
The proof is similar to the one concerning the leading terms of
$HU$ given in Lemma $III.1$ of \cite{param1}, being however more involved. 

The matrix whose determinant we must compute is obtained from $B$ by deleting the
lines and columns corresponding to the subsets $I_0$ and $J_0$ (as explained
in the previous section).
 
The matrix $B'_{\hat I_0,\hat J_0}$ has
\begin{itemize}
\item a line and column corresponding to $u_{\tilde \ell}$, the short variable of $\tilde \ell$
\item $n$ lines and columns corresponding to the $v$ variables of the $n-1$ tree lines of $\cal T$ and the supplementary line $\ell_2$
\item $n-1$ lines and columns associated to the hypermomenta.
\end{itemize}

We represent the determinant of this matrix by the Grassmann integral:
\beqa
\label{d11}
\det B'_{\hat I_0,\hat J_0}=\int  d \bar \psi^u_{\tilde \ell} 
 d \psi^u_{\tilde \ell}
d \bar \psi^v d \psi^v
d\bar \psi^p d\psi^p
 e^{-\bar \psi B'\psi}
\eeqa
The quadratic form in the Grassman variables in the above integral is:
\beqa
\label{eq:formQuad}
&&-\sum_V \sum_{l;i,j} \bar \psi^u_{\tilde \ell} (-1)^{i+1}\omega(i,j) \e^V_{\tilde \ell i} \e^V_{\ell j} \psi^v_{\ell}
\nonumber\\
&&-\sum_V \sum_{l;i,j} \bar \psi^v_{\ell} \omega(i,j)\e^V_{\ell i} \e^V_{\tilde \ell j}(-1)^{j+1} \psi^u_{\tilde \ell}
\nonumber\\
&& +\sum_V \bar \psi^u_{\tilde \ell} (-1)^{i+1}\e^V_{\tilde \ell i} \psi^{p_V}
-\sum_V \bar \psi^{p_V} (-1)^{i+1}\e^V_{\tilde \ell i} \psi^u_{\tilde \ell}
\nonumber\\
&&-\sum_{V} \sum_{\ell,\ell'; i,j} \bar \psi^v_{\ell} \omega(i,j) \e^V_{\ell i} \e^V_{\ell' j} \psi^v_{\ell'}
\nonumber\\
&&+ \sum_V \sum_{\ell; i} \bar \psi^v_{\ell} \e^V_{\ell i} \psi^{p_V} - \sum_V \sum_{\ell; i} \bar \psi^{p_V} \e^V_{\ell i} \psi^v_{\ell} \, .
\eeqa

We implement the first Filk move as a Grasmann change of variales. At each step we reduce a tree line $\ell_1=(i,j)$ connecting the root vertex $\bar{V}_G$ to a normal vertex $V$ and gluing the two vertices. This is achieved by performing a change of variables for the hypermomenta and reinterpreting the quadratic form in the new variables as corresponding to a new vertex $\bar V'_G$: the quadratic form essentially reproduces itself under the change of variables!

Take $\ell_1=(i,j)$ a line connecting the ''root" vertex $\bar V_G$ to a vertex $V$. We make the change of variables
\beqa
\label{eq:magic1}
\psi^{p_V}&=&\chi^{p_V}+\sum_{\ell' \neq \ell_1}\sum_{k}
\left(
-\omega(i,k)\e^{\bar V_G}_{\ell' k}+\omega(j,k)\e^{V}_{\ell' k}
\right)\psi^v_{\ell'} 
\nonumber\\
&&+\sum_{k}\left(
-\omega(i,k)\e^{\bar V_G}_{\tilde \ell k}+\omega(j,k)\e^{V}_{\tilde \ell k}
\right) (-1)^{k+1}\psi^u_{\tilde \ell}
\nonumber\\
\bar \psi^{p_V}&=&\bar \chi^{p_V}+\sum_{\ell' \neq \ell_1}\sum_{k}
\left(
-\omega(i,k)\e^{\bar V_G}_{\ell' k}+\omega(j,k)\e^{V}_{\ell' k}
\right)\bar \psi^v_{\ell'}
\nonumber\\
&&+\sum_{k}\left(
-\omega(i,k)\e^{\bar V_G}_{\tilde \ell k}+\omega(j,k)\e^{V}_{\tilde \ell k}
\right) (-1)^{k+1} \bar \psi^u_{\tilde \ell} \,,
\eeqa
where the first sum is performed on the internal lines of $G$ (note that because of
the presence of the incidences matrices $\e$ this sum reduces to a sum on the
lines hooked to the two vertices $\bar{V}_G$ and $V$). 
The corners $i$ and resp. $j$ are the corners where the tree line $\ell_1$ hooks to the vertex $\bar V_G$ and resp. $V$.  

At each step, let us consider the coupling between the variables associated to the line $\ell_1$ and to the hypermomentum $p_V$ and the rest of the variables.
Using eq. (\ref{eq:formQuad}) one  has
\beqa
\label{eq:magarie30}
&&-\bar \psi^u_{\tilde \ell}\sum_{p}(-1)^{p+1}
\left( \omega(p,i) \e^{V_G}_{\tilde \ell p} \e^{V_G}_{\ell_1 i} +
\omega(p,j)\e^{V}_{\tilde \ell p} \e^{V}_{\ell_1 j} \right)
\psi^v_{\ell_1}
\nonumber\\
&&-\bar \psi^v_{\ell_1} \sum_{p}
\left(\omega(i,p) \e^{V_G}_{\tilde \ell p} \e^{V_G}_{\ell_1 i} +
\omega(j,p)\e^{V}_{\tilde \ell p} \e^{V}_{\ell_1 j}
\right) \psi^u_{\tilde \ell}(-1)^{p+1}
\nonumber\\
&&+\sum_{V; p}\bar \psi^u_{\tilde \ell}(-1)^{p+1}\e^V_{\tilde \ell p} \psi^{p_V}
-\sum_{V; p}\bar \psi^{p_V} (-1)^{p+1}\e^V_{\tilde \ell p} \psi^u_{\tilde \ell}
\nonumber\\
&&-\bar \psi^v_{\ell_1}
\sum_{\ell'\neq \ell_1;k} 
\left(
\omega(i,k) \e^{\bar V_G}_{\ell_1 i} \e^{\bar V_G }_{\ell' k} 
+\omega(j,k) \e^{V}_{\ell_1 j} \e^{V}_{\ell' k}
\right)
\psi^v_{\ell'}
\nonumber\\
&&-\sum_{\ell'\neq \ell_1; k} \bar \psi^v_{\ell'} 
\left( \omega(k,i) \e^{\bar V_G}_{\ell' k} \e^{\bar V_G}_{\ell_1 i}
+ \omega(k,j) \e^{V}_{\ell' k} \e^{V}_{\ell_1 j}
\right) \psi^v_{\ell_1}
\nonumber\\
&&+\bar\psi^v_{\ell_1} \e^V_{\ell_1 j} \psi^{p_V}
+\sum_{\ell' \neq \ell_1; k} \bar \psi^v_{\ell'} \e^V_{\ell' k} \psi^{p_V}
\nonumber\\
&&-\bar\psi^{p_V} \e^V_{\ell_1 j} \psi^v_{\ell} -\sum_{\ell' \neq \ell_1; k} \bar \psi^{p_V} \e^V_{\ell' k} \psi^v_{\ell'} \, ,
\eeqa
which rewrites as
\beqa
\label{magarie2233}
&&-\Big{[}
\bar \psi^u_{\tilde \ell}\sum_{p}(-1)^{p+1}
\left( \omega(p,i) \e^{V_G}_{\tilde \ell p} \e^{V_G}_{\ell_1 i} +
\omega(p,j)\e^{V}_{\tilde \ell p} \e^{V}_{\ell_1 j} \right)
\nonumber\\
&&+\sum_{\ell'\neq \ell_1; k} \bar \psi^v_{\ell'} 
\left( \omega(k,i) \e^{\bar V_G}_{\ell' k} \e^{\bar V_G}_{\ell_1 i}
+ \omega(k,j) \e^{V}_{\ell' k} \e^{V}_{\ell_1 j}\right)
+\bar\psi^{p_V} \e^V_{\ell_1 j}
\Big{]}
\psi^v_{\ell_1}
\nonumber\\
&&-\bar \psi^v_{\ell_1} 
\Big{[}
\sum_{p}
\left(\omega(i,p) \e^{V_G}_{\tilde \ell p} \e^{V_G}_{\ell_1 i} +
\omega(j,p)\e^{V}_{\tilde \ell p} \e^{V}_{\ell_1 j}
\right) \psi^u_{\tilde \ell}(-1)^{p+1}
\nonumber\\
&&+\sum_{\ell'\neq \ell_1;k} 
\left(
\omega(i,k) \e^{\bar V_G}_{\ell_1 i} \e^{\bar V_G }_{\ell' k} 
+\omega(j,k) \e^{V}_{\ell_1 j} \e^{V}_{\ell' k}
\right)
\psi^v_{\ell'}
-\e^V_{\ell_1 j} \psi^{p_V}
\Big{]}
\nonumber\\
&&+\sum_{V; p}\bar \psi^u_{\tilde \ell}(-1)^{p+1}\e^V_{\tilde \ell p} \psi^{p_V}
-\sum_{V; p}\bar \psi^{p_V} (-1)^{p+1}\e^V_{\tilde \ell p} \psi^u_{\tilde \ell}
\nonumber\\
&&+\sum_{\ell' \neq \ell_1; k} \bar \psi^v_{\ell'} \e^V_{\ell' k} \psi^{p_V}
 -\sum_{\ell' \neq \ell_1; k} \bar \psi^{p_V} \e^V_{\ell' k} \psi^v_{\ell'} \, .
\eeqa
Performing now in \ref{eq:magarie30} the change of variable \ref{eq:magic1} 
for the hypermomentum $p_V$ of the vertex $V$ associated to the tree line $\ell_1$ of $G$ and taking into account that $\e_{\ell_1 i}=-\e_{\ell_1 j}$ the first two lines of \ref{magarie2233} are simply
\beqa
\label{eq:micamaga}
-\bar \chi^{p_V} \e^V_{\ell_1 j} \psi^v_{\ell_1}+
\bar \psi^v_{\ell_1} \e^V_{\ell_1 j} \chi^{p_V} \, .
\eeqa
As $\psi^v_{\ell_1}$ and $\bar \psi^v_{\ell_1}$ do not appear anymore in the rest of the terms we are forced to pair them with $\bar \chi^{p_V}$ and $\chi^{p_V}$. The rest of the terms in the quadratic form are:
\beqa
\label{eq:magaroi}
&&+\sum_{V; p}\bar \psi^u_{\tilde \ell}(-1)^{p+1}\e^V_{\tilde \ell p} 
\Big{[}
\sum_{\ell' \neq \ell_1}\sum_{k}
\left(
-\omega(i,k)\e^{\bar V_G}_{\ell' k}+\omega(j,k)\e^{V}_{\ell' k}
\right)\psi^v_{\ell'} 
\nonumber\\
&&+\sum_{k}\left(
-\omega(i,k)\e^{\bar V_G}_{\tilde \ell k}+\omega(j,k)\e^{V}_{\tilde \ell k}
\right) (-1)^{k+1}\psi^u_{\tilde \ell}
\Big{]}
\nonumber\\
&&+\sum_{\ell' \neq \ell_1; p} \bar \psi^v_{\ell'} \e^V_{\ell' p} 
\Big{[}
\sum_{\ell' \neq \ell_1}\sum_{k}
\left(
-\omega(i,k)\e^{\bar V_G}_{\ell' k}+\omega(j,k)\e^{V}_{\ell' k}
\right)\psi^v_{\ell'} 
\nonumber\\
&&+\sum_{k}\left(
-\omega(i,k)\e^{\bar V_G}_{\tilde \ell k}+\omega(j,k)\e^{V}_{\tilde \ell k}
\right) (-1)^{k+1}\psi^u_{\tilde \ell}
\Big{]}
\nonumber\\
&&-\sum_{V; p}
\Big{[}
\sum_{\ell' \neq \ell_1}\sum_{k}
\left(
-\omega(i,k)\e^{\bar V_G}_{\ell' k}+\omega(j,k)\e^{V}_{\ell' k}
\right)\bar \psi^v_{\ell'}
\nonumber\\
&&+\sum_{k}\left(
-\omega(i,k)\e^{\bar V_G}_{\tilde \ell k}+\omega(j,k)\e^{V}_{\tilde \ell k}
\right) (-1)^{k+1} \bar \psi^u_{\tilde \ell}
\Big{]}
(-1)^{p+1}\e^V_{\tilde \ell p} \psi^u_{\tilde \ell}
\nonumber\\
&& -
\Big{[}
\sum_{\ell' \neq \ell_1}\sum_{k}
\left(
-\omega(i,k)\e^{\bar V_G}_{\ell' k}+\omega(j,k)\e^{V}_{\ell' k}
\right)\bar \psi^v_{\ell'}
\nonumber\\
&&+\sum_{k}\left(
-\omega(i,k)\e^{\bar V_G}_{\tilde \ell k}+\omega(j,k)\e^{V}_{\tilde \ell k}
\right) (-1)^{k+1} \bar \psi^u_{\tilde \ell}
\Big{]}
\sum_{\ell' \neq \ell_1; p}
\e^V_{\ell' p} \psi^v_{\ell'}\; .
\eeqa

We analyse the different terms in the above equation.
The term $\bar \psi^u_{\tilde \ell} \psi^u_{\tilde \ell}$ is:
\beqa
&&\sum_{p,k}\e^V_{\tilde\ell p} (-1)^{p+1}\Big{[}
- \omega(i,k) \e^{\bar V_G}_{\tilde \ell k} + \omega(j,k) \e^{V}_{\tilde \ell k} 
\Big{]}(-1)^{k+1}
\nonumber\\
&&-\sum_{p,k}\Big{[}
- \omega(i,k) \e^{\bar V_G}_{\tilde \ell k} + \omega(j,k) \e^{V}_{\tilde \ell k} 
\Big{]}(-1)^{k+1}\e^V_{\tilde\ell p} (-1)^{p+1}=0 \, .
\eeqa

The term in $\bar \psi^u_{\tilde \ell} \psi^v_{\ell'}$ is given by:
\beqa
&&+\sum_{p}\bar \psi^u_{\tilde \ell}(-1)^{p+1}\e^V_{\tilde \ell p} 
\sum_{\ell' \neq \ell_1}\sum_{k}
\left(
-\omega(i,k)\e^{\bar V_G}_{\ell' k}+\omega(j,k)\e^{V}_{\ell' k}
\right)\psi^v_{\ell'}
\nonumber\\ 
&& -
\sum_{p}\left(
-\omega(i,p)\e^{\bar V_G}_{\tilde \ell p}+\omega(j,p)\e^{V}_{\tilde \ell p}
\right) (-1)^{p+1} \bar \psi^u_{\tilde \ell}
\sum_{\ell' \neq \ell_1; k}
\e^V_{\ell' k} \psi^v_{\ell'} \, .
\eeqa
Setting $j$ to be either the first or the last halfline on the vertex $V$ we see that the last terms in the two lines above cancel eachother. The first two terms hold:
\beqa
\sum_{p} \sum_{\ell' \neq \ell_1} \sum_{k} \bar \psi^u_{\tilde \ell} \psi^v_{\ell'}
(-1)^{p+1}\Big{[}
-\omega(i,k)\e^{\bar V_G}_{\ell' k}\e^V_{\tilde \ell p}
+\omega(i,p)\e^{\bar V_G}_{\tilde \ell p}\e^V_{\ell' k}
\Big{]} \, .
\eeqa
This can be rewritten in the form:
\beqa
-\sum_{p} \sum_{\ell' \neq \ell_1} \sum_{k} \bar \psi^u_{\tilde \ell} \psi^v_{\ell'}
(-1)^{p+1} \omega(p,k) \e_{\tilde \ell p}^{\bar V_G'}\e_{\ell' k}^{\bar V_G'} \, ,
\eeqa
where $\bar V_G'$ is a new root vertex in which the vertex $V$ has been glued to the vertex $\bar V_G$ and the halflines on the vertex $V$ have been inserted on the new vertex at the place of the halfline $i$.

The coupling between $\psi^v$'s with $\psi^v$'is
\beqa
\label{eq:maga333}
&+&\sum_{\ell\neq\ell_1,k}\bar \psi^v_{\ell} \e^{V}_{\ell k} 
 \sum_{\ell'}\left(-\sum_p \e^{\bar{V}_G}_{\ell' p} \omega(i,p)
+\sum_{k'=1}^4 \e^{V}_{\ell' k'} \omega(j,k')\right) \psi^v_{\ell'}
\nonumber\\
&-&\sum_{\ell' \neq \ell_1} \left( - \sum_p \e^{\bar{V}_G}_{\ell' p} \omega(i,p) + \sum_{k=1}^4 \e^{V}_{\ell' k} \omega(j,k)\right) \bar{\psi}^v_{\ell'}
 \sum_{\ell\neq\ell_1;k'}\e^{V}_{\ell k'}\psi^v_{\ell} \ .
\eeqa

Again, as $j$ is either the first or the last halfline on the vertex $V$, the last two terms in the two lines of eq. (\ref{eq:maga333}) cancel each other. The rest of the terms give exactly the contacts between the $\bar\psi^v$ and $\psi^v$ on a new vertex $\bar V_G'$ obtained by gluing $V$ on $\bar{V}_G$
This is the first Filk move on the line $\ell_1$ and its associated vertex $V$. 
One iterates now this mechanism for the rest of the tree lines of $G$. 
Hence we reduce the graph $G$ to a rosette (see subsection \ref{grafuri}). 

The quadratic form writes finally as
\beqa
&&-\sum_{\ell'; p,k} \bar \psi^u_{\tilde \ell} \psi^v_{\ell'}
(-1)^{p+1} \omega(p,k) \e_{\tilde \ell p}^{\bar V_G}\e_{\ell' k}^{\bar V_G}
-\sum_{\ell'; p,k }
\bar\psi^v_{\ell'} \psi^u_{\tilde \ell}(-1)^{p+1}
\omega(k,p) \e_{\tilde \ell p}^{\bar V_G}\e_{\ell' k}^{\bar V_G}
\nonumber\\
&&-\sum_{\ell, \ell';j,k} \bar \psi^v_\ell \e^{\bar {V}_G}_{\ell j} \e^{\bar {V}_G} _{\ell' k} \omega (j,k) \psi^v_{\ell'} \, .
\eeqa
The sum concerns only the rosette vertex $\bar {V}_G$. Therefore the last line is $0$, as by the first Filk move we have exhausted all the tree lines of $\cal T$.

As $\tilde \ell$ breaks the face separated from the external face by $\ell_2$ we have  
$k_1< p < k_2$. By a direct inspection of the terms above, one obtains the
requested result.

\qed

\section{Factorization of the Feynman amplitudes}
\label{sec:factorisation}
\setcounter{equation}{0}

Take now $S$ to be a primitively divergent subgraph.
We now prove that the Feynman amplitude ${\cal A}_G$ factorizes into two
parts, one corresponding to the primitive divergent subgraph $S$ and the other to the graph $G/S$ (defined in section \ref{grafuri}). This is needed in order to prove that divergencies are cured by Moyal counterterms
\footnote{
 It is also the property needed for the definition of a coproduct $\Delta$ for a Connes-Kreimer Hopf algebra structure (see \cite{ck1,ck2, kreimer}). Details of this construction are given elsewhere \cite{progres}.}.

In section \ref{sec:NCdimreg} we will prove that only the planar ($g=0$), one broken face $B=1$, $N=2$ or $N=4$ external legs subgraphs are primitively divergent. We now deal only with such subgraphs.

\subsection{Factorization of the polynomial $HU$}

We denote all the leading terms in the first polynomial associated to a graph $S$ by $HU^{l}_S$. If we rescale all the $t_{\ell}$ parameters corresponding to a subgraph $S$ by $\rho^2$, $HU_G$ becomes a polynomial in $\rho$. We denote the terms of minimal degree in $\rho$ in this polynomial by $HU_G^{l( \rho)}$. It is easy to see that for the subgraph $S$ we have 
$HU^{l(\rho)}_{S}=\rho^{2[L(S)-n(S)+1]}HU^{l}_{S}\vert_{\rho=1}$.
We have the following theorem
\begin{theorem}
\label{mare}
Under the rescaling
\beqa
\label{rescalare}
t_{\alpha}\mapsto \rho^2 t_{\alpha}
\eeqa
of the parameter corresponding to a divergent subgraph $S$ of any Feynman
graph $G$, the following factorization property holds
\beqa
HU^{l(\rho)}_{G,\bar V}=HU^{l(\rho)}_{S,\bar V_S} HU_{G/S,\bar V} \; .
\eeqa
\end{theorem}
{\bf Proof:} \\ 
In the matrix $M$ defined in eq. (\ref{M}) (corresponding to the graph $G$) we can rearrange the lines and columns so that we  place the matrix $M_S$ (corresponding to the subgraph $S$) into the upper left corner.
We place the line (and resp. the column) associated to the hypermomentum
of the root vertex of $S$ to be the last line (and resp. column) of $M$ 
(without loss of generality we consider that the root of the subgraph $S$ is not
the root of $G$). $M$ takes the form
\beqa
\label{M1}
\hskip -.2cm
\begin{pmatrix} E^{u^Su^S} & E^{u^Sv^S} & C^{u^Sp^S}  & E^{u^Su^{G-S}}&
  E^{u^Sv^{G-S}} & C^{u^Sp^{G-S}} \\
 E^{v^Su^S} & E^{v^Sv^S} & C^{v^Sp^S}  & E^{v^Su^{G-S}}&
  E^{v^Sv^{G-S}} & C^{v^Sp^{G-S}} \\
 C^{p^Su^S} & C^{p^Sv^S} & 0  & C^{p^Su^{G-S}} &
  C^{p^Sv^{G-S}} & 0\\
 E^{u^{G-S}u^S} & E^{u^{G-S}v^S} & C^{u^{G-S}p^S}  & E^{u^{G-S}u^{G-S}}&
  E^{u^{G-S}v^{G-S}} & C^{u^{G-S}p^{G-S}}\\
 E^{v^{G-S}u^S} & E^{v^{G-S}v^S} & C^{v^{G-S}p^S}  & E^{v^{G-S}u^{G-S}}&
  E^{v^{G-S}v^{G-S}} & C^{v^{G-S}p^{G-S}}\\
 C^{p^{G-S}u^S} & C^{p^{G-S}v^S} & 0  & C^{p^{G-S}u^{G-S}}&
  C^{p^{G-S}v^{G-S}} & 0
\end{pmatrix}\, ,
\eeqa
where we have denoted by $E^{u^Su^S}$ a  coupling between two short
variables corresponding to internal lines of $S$ {\it etc.}.

{\bf I.} 
We first write the determinant of the matrix above under the form of a Grassmannian
integral 
\beqa
\label{d1}
\det M=\int  d \bar \psi^u d \psi^u  d \bar \psi^v d \psi^v d\bar \psi^p d\psi^p
 e^{- \bar \psi M \psi} \, .
\eeqa

Denote a generic line of the subgraph $S$ by $\ell_S$ and a generig line of the subgraph $G-S$ by $\ell_{G-S}$.

We perform a Grassmann change of variables of Jacobian $1$. The value of the integral (\ref{d1}) does not change under this change of variables. We will prove that the following properties hold for the different terms in the Grassmanian quadratic form
\beqa
\label{eq:scop}
&&E'^{v^Sv^S}=\mathrm{diag}(t_l), \quad E'^{v^Su^{G-S}}=0,
\quad E'^{v^Sv^{G-S}}=0\nonumber\\
&&E'^{u^{G-S}u^{G-S}}=E'^{u^{G/S}u^{G/S}}, \quad E'^{u^{G-S}v^{G-S}}=E'^{u^{G/S}v^{G/S}} \nonumber\\
&&E'^{v^{G-S}v^{G-S}}=E'^{v^{G/S}v^{G/S}}.
\eeqa
The new matrix of the quadratic form $M'$will now be 
\beqa
\label{M2}
\hskip -.2cm
\begin{pmatrix} E'^{u^Su^S} & E'^{u^Sv^S} & C^{u^Sp^S}  & E'^{u^Su^{G-S}}&
  E'^{u^Sv^{G-S}} & C^{u^Sp^{G-S}} \\
 E'^{v^Su^S} & E'^{v^Sv^S} & C^{v^Sp^S}  & 0 &
  0  & C^{v^Sp^{G-S}} \\
 C^{p^Su^S} & C^{p^Sv^S} & 0  & C^{p^Su^{G-S}} &
  C^{p^Sv^{G-S}} & 0\\
 E'^{u^{G-S}u^S} & 0 & C^{u^{G-S}p^S}  & E'^{u^{G/S}u^{G/S}}&
  E'^{u^{G/S}v^{G/S}} & C^{u^{G-S}p^{G-S}}\\
 E'^{v^{G-S}u^S} & 0  & C^{v^{G-S}p^S}  & E'^{v^{G/S}u^{G/S}}&
  E'^{v^{G/S}v^{G/S}} & C^{v^{G-S}p^{G-S}}\\
 C^{p^{G-S}u^S} & C^{p^{G-S}v^S} & 0  & C^{p^{G-S}u^{G-S}}&
  C^{p^{G-S}v^{G-S}} & 0
\end{pmatrix}\, .
\eeqa

The first part of the proof follows that of section \ref{sec:furtherLeading}, where we replace 
\beqa
\label{eq:improuvement}
 \psi^u_{\tilde \ell} \rightarrow \sum_{\ell'\in G-S} \psi^u_{\ell'}
 \qquad
  \bar \psi^{u}_{\tilde \ell} \rightarrow \sum_{\ell'\in G-S} \bar \psi^{u}_{\ell'} \, .
\eeqa
Thus the appropriate change of variables is now
\beqa
\label{eq:magic1'}
\psi^{p_V}&=&\chi^{p_V}+\sum_{\ell' \neq \ell_1; k}
\left(
-\omega(i,k)\e^{\bar V_S}_{\ell' k}+\omega(j,k)\e^{V}_{\ell'; k}
\right)\psi^v_{\ell'} 
\nonumber\\
&&+\sum_{\genfrac{}{}{0pt}{}{\ell'; k }{\ell'\in G-S}}\left(
-\omega(i,k)\e^{\bar V_S}_{\ell' k}+\omega(j,k)\e^{V}_{ \ell' k}
\right) (-1)^{k+1}\psi^u_{\ell'}
\nonumber\\
\bar \psi^{p_V}&=&\bar \chi^{p_V}+\sum_{\ell' \neq \ell_1; k}
\left(
-\omega(i,k)\e^{\bar V_S}_{\ell' k}+\omega(j,k)\e^{V}_{\ell' k}
\right)\bar \psi^v_{\ell'}
\nonumber\\
&&+\sum_{\genfrac{}{}{0pt}{}{\ell'; k }{\ell'\in G-S}}\left(
-\omega(i,k) \e^{\bar V_S}_{ \ell' k} + \omega(j,k)\e^{V}_{ \ell' k}
\right) (-1)^{k+1} \bar \psi^u_{ \ell'} \; .
\eeqa

We emphasize that this Grassmann change of variables can be viewed as forming appropriate linear combinations of lines and columns. As we only use lines and columns associated to hypermomenta $p_S$, this manipulations can not change the value of the determinant in the upper left corner: it will allways correspond precisely to the first polynomial of the subgraph $S$.

The relevant terms in the quadratic form are those of  eq. (\ref{eq:magarie30}) 
and eq. (\ref{magarie2233}), with the substitutions (\ref{eq:improuvement}).
Again, after the change of variables \ref{eq:magic1'} the only surviving contacts of $\psi^v_{\ell_1}$ and $\bar \psi^v_{\ell_1}$ are given by 
eq. (\ref{eq:micamaga}).
Finally, the remaining terms are given by  eq. (\ref{eq:magaroi}) with the substitutions (\ref{eq:improuvement}).

One needs again to analyse the different terms in this equation.
The quadratic term in $\bar\psi^u \psi^u$ is:
\beqa
&&\sum_{\ell',\ell''\in G-S}\bar \psi ^u_{\ell'}\psi ^u_{\ell''}(-1)^{k+p+1}
\Big{(}
\omega(i,k)\e^{V}_{\ell' p}\e^{\bar V_S}_{\ell'' k}-
\omega(j,k)\e^{V}_{\ell' p}\e^{\bar V}_{\ell'' k}
\nonumber\\
&&-\omega(i,k)\e^{\bar V_S}_{\ell' k}\e^{V}_{\ell'' p}
+\omega(j,k)\e^{V}_{\ell' k}\e^{V}_{\ell'' p}
\Big{)}\, .
\eeqa
As $j$ is either the first or the last halfline on the vertex $V$, we see that the last terms in the two lines above cancel. The remaining two terms give the contacts amongst $\bar\psi^u$ and $\psi^u$ on the rosette new vertex $ \bar V'_S$, obtained by gluing $\bar V_S$ and $V$ respecting the ordering.

The contacts between $\bar\psi^u$ and $\psi^v$ become
\beqa
&&+\sum_{\ell'}\sum_{p}\bar \psi^u_{\ell'}(-1)^{p+1}\e^V_{\ell' p} 
\sum_{\ell'' \neq \ell_1}\sum_{k}
\left(
-\omega(i,k)\e^{\bar V_G}_{\ell'' k}+\omega(j,k)\e^{V}_{\ell'' k}
\right)\psi^v_{\ell''}
\nonumber\\ 
&& -
\sum_{\ell'}\sum_{p}\left(
-\omega(i,p)\e^{\bar V_G}_{ \ell' p}+\omega(j,p)\e^{V}_{ \ell' p}
\right) (-1)^{p+1} \bar \psi^u_{\ell'}
\sum_{\ell'' \neq \ell_1; k}
\e^V_{\ell'' k} \psi^v_{\ell''} \, .
\eeqa 
Again that the last terms in the two lines above cancel. Rearanging the rest as before we end recover again the terms corresponding to a rosette.

Finally for the $\bar\psi^v$ and $\psi^v$ contacts eq .(\ref{eq:maga333})  goes through.

We iterate the change of variables {\bf only} for a tree in the subgraph $S$, hence we reduce the subgraph $S$ to a rosette (see subsection \ref{grafuri}). As the quadratic form reproduced itself, the root vertex $\bar V_S$ is now a Moyal vertex, with either $2$ or $4$ external legs.

Let $r$ be an external half line of the subgraph $S$. As $S$ is planar one broken face any line $l_S=(p,q)$ will either have
$r<p,q$ or $p,q<r$. Thus
\beqa
E'^{v^Sv^{G-S}}=\sum_{l,p,r}\omega(p,r)\e^{\bar V_S}_{\ell_S p}\e^{\bar V_S}_{ \ell_{G-S} r}=0 \,.
\eeqa
The reader can check by similar computations that eq. (\ref{eq:scop}) holds. 

\medskip

\noi{\bf II.} To obtain in the  lower right corner the matrix corresponding
to the graph $G/S$, we just have to add the lines and
columns of the hypermomenta corresponding to the vertices of $S$ to the ones
corresponding to the root of $S$. Furthermore, by performing this operation,
the block $ C^{v^Sp^{G-S}}$ (which had only one non-trivial column, the column
corresponding to the hypermomentum $p_{\bar V_S}$) becomes identically $0$. Forgetting the primes, the matrix in the quadratic becomes:
\beqa
\label{M3}
\begin{pmatrix} E^{u^Su^S} & E^{u^Sv^S} & C^{u^Sp^S}  & E^{u^Su^{G-S}}&
  E^{u^Sv^{G-S}} & C^{u^Sp^{G-S}} \\
 E^{v^Su^S} & t_{\ell} \delta_{v^Sv^S} &  C^{v^Sp^S} & 0 &
  0  & 0\\
 C^{p^Su^S} & C^{p^Sv^S} & 0  & C^{p^Su^{G-S}} &
  C^{p^Sv^{G-S}} & 0\\
 E^{u^{G-S}u^S} & 0 & C^{u^{G-S}p^S}  & E^{u^{G/S}u^{G/S}}&
  E^{u^{G/S}v^{G/S}} & C^{u^{G/S}p^{G/S}}\\
 E^{v^{G-S}u^S} & 0  & C^{v^{G-S}p^S}  & E^{v^{G/S}u^{G/S}}&
  E^{v^{G/S}v^{G/S}} & C^{v^{G/S}p^{G/S}}\\
 C^{p^{G-S}u^S} & 0 & 0  & C^{p^{G/S}u^{G/S}}&
  C^{p^{G/S}v^{G/S}} & 0
\end{pmatrix}.
\eeqa
This is equivalent to the Grassmannian change of variables:
\beqa
\chi'^{p^S}&=&\chi^{p^S}+\psi^{p_{\bar V_S}}\, ,\nonumber\\
\bar \chi'^{p^S}&=&\bar \chi^{p^S}+\bar \psi^{p_{\bar V_S}}\, .
\eeqa

\medskip

\noi
{\bf III.} We finally proceed with the rescaling 
 with $\rho^2$ of all the parameter $t_\alpha$ coresponding
to the divergent subgraph $S$ (see \ref{rescalare}). Recall that these parameters are present as
$\frac{1}{t_\alpha}$ on the diagonal of the block $E^{u^Su^S}$ and as
$t_\alpha$  on the diagonal of the block $E^{v^Sv^S}$.
 
We factorize $\frac{1}{\rho}$ on the first $L(S)$ lines of
columns of $M$ (corresponding to the $u^S$ variables) and $\rho$ on the next $L(S)$ lines and columns (coresponding to the $v^S$ variables). 
We also factorize $\frac{1}{\rho}$ on the $n(S)-1$ lines and columns
corresponding to the hypermomenta $p^S$. This is given by the Grassmann change of variables
\beqa
\label{ultima}
\psi'^u&=&\frac{1}{\rho}\psi^u\nonumber\\
\psi'^v&=&{\rho}\psi^v\nonumber\\
\chi'^p&=&\frac{1}{\rho}\chi^p\, .
\eeqa

In the new variables the matrix of the quadratic form is
\beqa
\label{M4}
\hskip -.4cm
\begin{pmatrix} c_l\delta_{uu'}+\rho^2 E'^{uu'} & E'^{uv} & \rho^2 C^{up}  & \rho E'^{uu}&
 \rho E'^{uv} & \rho C'^{up} \\
 E'^{vu} & t_l\delta_{vv'} & C^{vp}  & 0 & 0  & 0 \\
 \rho ^2 C^{pu} & C^{pv} & 0  & \rho C^{pu} & \rho C^{pv} & 0\\
 \rho E'^{uu} & 0 & \rho C^{up}  & E'^{u^{G/S}u^{G/S}}&
  E'^{u^{G/S}v^{G/S}} & C'^{u^{G/S}p^{G/S}}\\
 \rho E'^{vu} & 0  & \rho C^{vp}  & E'^{v^{G/S}u^{G/S}}&
  E'^{v^{G/S}v^{G/S}} & C'^{v^{G/S}p^{G/S}}\\
 \rho C'^{pu} & 0 & 0  & C'^{p^{G/S}u^{G/S}}& C'^{p^{G/S}v^{G/S}} & 0
\end{pmatrix} .
\eeqa

The determinant of the original matrix is obtained by multiplying the overall factor $\rho^{-2n(S)+2}$ (coming from the Jacobian of the change of variables) with the determinant of the matrix (\ref{M4}). To obtain $HU_G$, according to eq. (\ref{hugvq2}) we must multiply this determinant by a product over all lines of $t_{\ell}$.

The determinant upper left corner, coresponding to the subgraph $S$, multiplied by the appropriate product of $t_{\ell}$ as in eq. (\ref{hugvq2}) and by the Jacobian factor holds the complete polynomial $HU_S$. At leading order in $\rho$ it is $HU_S^{l(\rho)}$ 

The determinant of the lower right corner, multiplied by its corresponding product of $t_{\ell}$ holds the complete polynomial $HU_{G/S}$ and no factor $\rho$.

At the leading order in $\rho$ the off diagonal blocks become $0$. Therefore we have
\beqa
  HU^{l(\rho)}_G=HU^{l(\rho)}_S HU_{G/S} \, .
\eeqa

\qed

Let us illustrate all this with the example of the graph of Fig. \ref{hyper}, where the primitive divergent subgraph is taken to be the sunshine graph of lines
$\ell_4,\ell_5$ and $\ell_6$. A direct computation showed that, under the
rescaling
\beqa
t_4\to \rho^2 t_4,\ t_5\to \rho^2 t_5,\ t_6\to \rho^2 t_6,\
\eeqa
the leading terms in $\rho$ of the polynomial $HU_G$ factorize as
\beqa
\rho^{4}\left( (1+4s^2) t_4 (t_5+t_6) +t_5 (t_6+8s^2(2 t_5 +t_6+ 2 s^2
  t_6)\right)\nonumber\\
(1+4s^2) (t_1 + t_2 + t_3 + t_1 t_2 t_3) (1+ t_2 t_3 + t_1 (t_2+ t_3)).
\eeqa
The first line of this,formula corresponds to the leading terms under the
rescaling with $\rho$ of $HU_S$, while the second line is nothing but the
polynomial $HU_{G/S}$ of eq. (\ref{bula-traficata}).

\subsection{The exponential part of the Feynman amplitude}

In order to perform the appropriate subtractions we need to check the factorization also at the level of the second polynomial.
Throughout this section we suppose that $S$ is a completely internal subgraph, that is none of its external points is an external point of $G$. The general case is treated by the same methods with only slight modifications.
We have the following lemma
\begin{proposition}
   Under the rescaling $t_{\alpha}\mapsto \rho^2 t_{\alpha}$ of all the lines of the subgraph $S$ we have:
   \beqa
    \frac{HV_G}{HU_G}\Big{\vert}_{\rho =0}=\frac{HV_{G/S}}{HU_{G/S}}\, .
   \eeqa
\end{proposition}
{\bf Proof:} The ratio $\frac{HV_G}{HU_G}$ is given by the inverse matrix $Q^{-1}$ due to eq. (\ref{eq:HVGV}) which in turn is given by $M^{-1}$ (see \cite{param1} for the exact relation). Thus any property which holds for $M^{-1}$ will also hold for $Q^{-1}$.

We write the matrix elements of the inverse of $M$ with the help of Grassmann variables
\beqa
(M^{-1})_{ij}=\frac{\int d\bar \psi d\psi \psi_i \bar \psi_j e^{-\bar \psi M \psi}}{\int d\bar \psi d\psi  e^{-\bar \psi M \psi}}
\, .
\eeqa

As $S$ is a completely internal subgraph we only must analyse the inverse matrix entries
\beqa
(M^{-1})_{G-S G-S}=\frac{\int d\bar \psi d\psi \psi_{G-S} \bar \psi_{G-S} e^{-\bar \psi M \psi}}{\int d\bar \psi d\psi  e^{-\bar \psi M \psi}} \, ,
\eeqa
the only ones which intervene in the quadratic form due to the matrix $P$ in \ref{eq:HVGV}.
None of the changes of variables of the previous section involve any Grassmann variable associated with the $G-S$ sector. We conclude that
\beqa 
(M^{-1})_{G-S G-S}=(M'^{-1})_{G-S G-S} \, ,
\eeqa
with $M'$ in (\ref{M3}). After the rescaling with $\rho$, the matrix $M'$ becomes
(\ref{M4}). At leading order we set $\rho$ to zero so that $M'$ becomes block diagonal. Consequently
\beqa
M^{-1}_{G-S G-S}=M'^{-1}_{G/SG/S} \, .
\eeqa

\qed

\subsection{The two point function}
\label{sec:2point}

The results proven above must be refined further for the two point function. The reason is that, as explained in section \ref{sec:NCdimreg}, the two point functions have two singularities so that one needs also to analyse subleading behaviour.

In the sequel we replace $Q_G$ by $M_G$, $Q_{G/S}$ by $M_{G/S}$, etc., the difference between the $Q$'s and the $M$'s being inessential.

When integrating over the internal variables of $G$ we start by integrating over the variables associated to $S$ first. All the variables $u$, $v$ and $p$ appearing in the sequel belong then to $G/S$.
The amplitude of the graph $G$ will then write, after rescaling of the parameters of the subgraph $S$ and having perform the first Filk move
\beqa
{\cal A_G}=\int [ d t_\ell d\rho] \int [dudvdp]^{G/S}
\rho^{2L(S)-1}\frac
{e^{- \begin{pmatrix}
       u & v & p 
      \end{pmatrix}
(M'_{G/S}+\rho^2 \delta M')
       \begin{pmatrix}
         u \\ v \\ p
      \end{pmatrix}
          }}
{HU_{S}^{\frac D2}(\rho)},
\eeqa
where $[dt d\rho]$ is a short hand notation for the measure of integration on the Schwinger parameters, to be developped further in the next section.
$[dudvdp]^{G/S}$ is the measure of integration for the internal variables of $G/S$, and $M'$ is given in eq. (\ref{M4}). We explicitate $\delta M'$ as
 \beqa
\label{eq:deltaM'}
\delta M'=&& 
  \begin{pmatrix}
 E'^{u^{G-S}u^S} & 0 & C^{u^{G-S}p^S} \\
  E'^{v^{G-S}u^S} & 0  & C^{v^{G-S}p^S}\\
 C'^{p^{G-S}u^S} & 0 & 0
  \end{pmatrix}
  \begin{pmatrix}
c_l\delta_{u^Su'^S} & E'^{u^Sv^S} & 0 \\
 E'^{v^Su^S} & t_l\delta_{v^Sv'^S} & C^{v^Sp^S} \\
 0 & C^{p^Sv^S} & 0
  \end{pmatrix}^{-1}
\nonumber\\
&&
  \begin{pmatrix}
  E'^{u^Su^{G-S}}& E'^{u^Sv^{G-S}} & C'^{u^Sp^{G-S}} \\
   0 & 0  & 0 \\
   C^{p^Su^{G-S}} & C^{p^Sv^{G-S}} & 0\\
  \end{pmatrix} \, .
\eeqa

The Taylor development in $\rho$ of the exponential gives
\beqa
\label{eq:terms}
&&\int [ d t_\ell] \int [dudvdp]^{G/S}
{e^{- \begin{pmatrix}
       u & v & p 
      \end{pmatrix}
        M'_{G/S}
       \begin{pmatrix}
         u \\ v \\ p
      \end{pmatrix}
          }}\nonumber\\
&&\int d\rho \rho^{2L(S)-1}(-1)\frac
{1+ \rho^2\begin{pmatrix}
       u & v & p 
      \end{pmatrix}
       \delta M'
       \begin{pmatrix}
         u \\ v \\ p
      \end{pmatrix}
          }
{HU_{S}^{\frac D2}(\rho)} \, .
\eeqa

The first term in the integral over $\rho$ above corresponds to a (quadratic) mass divergence.

The second term (logarithmically divergent) coresponds to the insertion of some operator which we now to compute. 

The interaction is real. This means that we should symmetrize our amplitudes over complex conjugation of all 
vertices. For instance, at on loop one should allways symmetrize the left and right tadpoles \cite{landau1,landau2,landau3}
\footnote{Following \cite{4men} one can prove using this argument that if terms like $x \partial$ do not appear in the initial 
lagrangean for the complex orientable model, they will not be generated by radiative corrections, which not proven there.}. 

Consequently, the inverse matrix in eq. (\ref{eq:deltaM'}) is actually a sum over the two possible choices of orientation of vertices. We must also sum over all possible choices of signs for the entries in the contact matrices in eq. (\ref{eq:deltaM'}) as a similar symmetrization must be performed for the hypermomenta.

As $\bar \psi^{u^S}$ couples only to the linear combination $\psi^{v^{G-S}}_{\ell}+\e^{V}_{\ell i}\psi^{u^{G-S}}_{\ell}$ in the initial matrix as well as in the change of variables, we have $E'^{u^Su^{G-S}}=\e^V_{l_{G-S}i} E'^{u^Sv^{G-S}}$.

Due to the sums over choices of signs, the only non zero entries in $\delta M'$ are $\delta M'_{u^S u^S}$, $\delta M'_{v^S v^S}$, $\delta M'_{v^S u^S}$, $\delta M'_{u^S v^S}$ and $\delta M'_{p_{\bar V_S} p_{\bar V_S}}$. 

We denote the two external lines of $S$ by $\ell_1$ and $\ell_2$. A tedious but straightforeward computation holds
\beqa
\begin{pmatrix}
       u & v & p 
      \end{pmatrix}
       \delta M'
       \begin{pmatrix}
         u \\ v \\ p
      \end{pmatrix}
=\Big{(}
A_1(\e^{\bar V_S}_{\ell_1 i}u_{\ell_1}+v_{\ell_1})^2+
A_2(\e^{V}_{\ell_2 i}u_{\ell_2}+v_{\ell_2})^2+
B_1 p_{\bar V_S}^2
\Big{)} .
\eeqa

The first two terms are an insertion of the operator $\Omega x^2$ whereas the last is the insertion of an operator $-\Delta+x^2$, being of the form of the initial lagrangean.

Take the graph $G/S$ with the insertion of this operator at $S$, and with the adition of eventual mass subdivergencies (due to the first term). We denote it by an operator ${\cal O}_S$ action on the graph $G/S$. 

We can then sum up the results of this section in the formula:
\beqa
\label{eq:factfinal}
\frac{e^{-\frac{HV_G(\rho)}{HU_G(\rho)}}}{HU_G(\rho)^{D/2}}=
\frac{1}{[HU^{l(\rho)}_{S}]^{D/2}}(1+\rho^2{\cal O}_S) \frac{e^{-\frac{HV_{G/S}}{HU_{G/S}}}}{HU_{G/S}^{D/2}} \, .
\eeqa

\section{Dimensional regularization and renormalization of NCQFT}
\label{sec:NCdimreg}
\setcounter{equation}{0}

In this final section we proceed to the dimensional regularization and renormalization of NCQFT. We detail the meromorphic structure and give the form of the subtraction operator. Dimensional regularization and meromorphie of Feynam amplitudes for this model was also established in \cite{ncmellin}. However, for consistency reasons we will give here an independent proof of this results.

However, as the proof of convergence of the renormalized integral for this $\Phi^{\star 4}_4$ model is identical with that for the commutative $\Phi^4_4$ (up to substituting the commutative subtraction operator with our subtraction operator) we will not detail it here.

\subsection{Meromorphic structure of NCQFT}

In this subsection we prove the meromorphic structure of a Feynman amplitude ${\cal A}$. We follow here the approach of \cite{reg}. We express the amplitude by eq. (\ref{HUGV}) 
\beqa
\label{HUGV-t}
{\cal A}_{G,{\bar V}}  (x_e,\;  p_{\bar V}, D) = \left(\frac{\tom}{2^{\frac
    D2 -1}}\right)^L  \int_{0}^{1} \prod_{\ell=1}^L  
 dt_\ell (1-t_\ell^2)^{\frac D2 -1} 
\frac{
e^{-  \frac {HV_{G, \bar{V}} ( t_\ell , x_e , p_{\bar v})}{HU_{G, \bar{v}} ( t )}}
}
{HU_{G, \bar{V}} ( t )^{D/2}} \, .
\eeqa
We restrict our analysis to connected non-vacuum graphs. 
As in the commutative case we extend this expression to the entire complex plane. Take a Hepp sector $\sigma$ defined as 
\beqa
\label{hepp-nc}
0\le t_1 \le \ldots \le t_L \, ,
\eeqa
and perform the change of variables
\beqa
\label{change-nc}
t_\ell=\prod_{j=\ell}^L x_j^2,\ \ell =1,\ldots, L.
\eeqa
We denote by $G_i$ the subgraph composed by the lines $t_1$ to $t_i$. As before, we denote $L(G_i)=i$ the number of lines of $G_i$, $g(G_i)$ its genus, $F(G_i)$ its number of faces, etc..
The amplitude is
\beqa
{\cal A}_{G,\bar V}=&&\Big{(}\frac{\tilde \Omega}{2^{(D-4)/2}}\Big{)}^L
\int_{0}^{1}\prod_{i=1}^L 
\left(1-(\prod_{j=i}^L x_j^2)^2\right)^{\frac D2 -1}
dx_{i} 
\nonumber\\
&&\prod_{i=1}^{L}x_{i}^{2L(G_i)-1}
\frac{e^{-\frac{HV_{G,\bar V}(x^2)}{HU_{G,\bar V}(x^2)}}}{HU_{G,\bar V}(x^2)}\, .
\eeqa 
In the above equation we factor out in $HU_{G,\bar V}$the monomial with the smallest degree in each variable $x_i$
\beqa
\label{ampli-x}
{\cal A}_{G,{\bar V}}  (x_e,\;  p_{\bar v}) = &&\left(\frac{\tom}{2^\frac
    D2}\right)^L  \int_{0}^{1} \prod_{\ell=1}^L  
 dx_\ell \left(1-(\prod_{j=\ell}^L x_j^2)^2\right)^{\frac D2 -1} 
\nonumber\\ 
&& x_i^{2L(G_i)-1-D b'(G_i)}
 \frac{e^{-\frac {HV_{G, \bar V}}{HU_{G,\bar V}}}}{(a s^b+ F (x^2))^\frac D2}.
\eeqa
The last term in the above equation is always bounded by a constant. Divergences can arise only in the region $x_i$ close to zero (it is well known that this theory does not have an infrared problem, even at zero mass).

The integer $b'(G_i)$ is given by the topology of $G_i$. It is
\beqa
b'(G_i)=\begin{cases}
    {\displaystyle \le L(G_i)-[n(G_i)-1]-2g(G_i)} &\text{if } 
      g(G_i)>0
     \vspace{.3cm}\\
     {\displaystyle \le L(G_i)-n(G_i)} &\text{if }
      g(G_i)=0 \text{ and } B(G_i)>1
     \vspace{.3cm}\\
     {\displaystyle =L(G_i)-[n(G_i)-1]} &\text{if }
       g(G_i)=0 \text{ and } B(G_i)=1 \\
  \end{cases} .
\eeqa

To prove the first and the third line one must look at the scaling of a leading term with $I=\{1 \hdots L \}$ and $J$ admissible in $HU_G$. 
For the second line one must take $I=\{1 \hdots L \}-\tilde\ell$ 
and $J$ {\it pseudo-admissible}. We have proved that such terms exists in section 
\ref{sec:furtherLeading}.

We see that $b'(G_i)$ is at most $L(G_i)-n(G_i)+1$ and that the maximum is achieved if and only if $g(G_i)=0$ and $B(G_i)=1$.

The convergence in the UV regime ($x_i\to 0$) is ensured if
\beqa
 \Re [ 2L(G_i)-Db'(G_i) ]>0 ,\ i=1\ldots L \, .
\eeqa
As
\beqa
\Re[2L(G_i)-Db'(G_i)]>\Re\Big{(}2L(G_i)-D[L(G_i)-n(G_i)+1]\Big{)} \, ,
\eeqa
we always have convergence provided
\beqa
\Re D<2\le \frac{4n(G_i)-N(G_i)}{n(G_i)-N(G_i)/2+1} \le \frac{2L(G_i)}{L(G_i)-n(G_i)+1} \, .
\eeqa
where $N(G_i)$ is the number of external points of $G_i$
\footnote{We have used here the topological relation 
$4n(G_i)-N(G_i)=2L(G_i)$}.
Thus ${\cal A}_{G, \bar V}(D)$ is analytic in the strip
\beqa
\label{domain-nc}
{\cal D}^\sigma=\{ D\, | \, 0 < \Re\, D <  2 \}.
\eeqa

We extend now the ${\cal A}$ as a function of $D$ for $2 \le \Re D \le 4$. We claim that if
\begin{itemize}
\item $g(G_i)>0$
\item $g(G_i)=0$ and $B(G_i)>1$
\item $N(G_i)>4$ \, ,
\end{itemize}
the strip of analyticity can be immediately extended up to 
\beqa
{\cal D}^\sigma=\{ D\, | \, 0 < \Re\, D <  4+\e_G \}.
\eeqa
for some small positive number $\e_G$ depending on the graph.
Indeed, for the first two cases we have $b'(G_i)\le L(G_i)-n(G_i)$ so that the integral over $x_i$ converges for
\beqa
\Re D\le 4 < \frac{4n(G_i)-N(G_i)}{n(G_i)-N(G_i)/2}=\frac{2L(G_i)}{L(G_i)-n(G_i)}
\, ,
\eeqa
whereas in the third case, as $N(G_i)>4$ the integral over $x_i$ converges for
\beqa
\Re D\le 4 < \frac{4n(G_i)-N(G_i)}{n(G_i)-N(G_i)/2+1}=
\frac{2L(G_i)}{L(G_i)-n(G_i)+1} \, .
\eeqa

The only possible divergences in ${\cal A}_{G,\bar V}(D)$ are generated by planar one two or four external legs subgraphs with a single broken face. They are called primitively divergent subgraphs.

Let $S$ be a primitively divergent subgraph and call $\rho$ its associated Hepp parameter. Using eq. \ref{eq:factfinal} its contribution to the amplitude writes:
\beqa
\label{eq:polamp}
{\cal A}^{\rho}_{G,\bar V_G}\sim
\int_{0}d\rho \rho^{2L(S)-1-D[L(S)-n(S)+1]}
\frac{1}{HU^l_{S,\bar V_{S}}|_{\rho=1}}(1+\rho^2 {\cal O}_S)
\frac{e^{-\frac{HV_{G/S}}{HU_{G/S}}}}{HU_{G/S}}\, .
\eeqa

The integral over $\rho$ is a meromorphic operator in $D$ with the divergent part given by
\beqa
h(D)=\frac{r_1}{2L(S)-D[L(S)-n(S)+1]}+
     \frac{r_2}{2L(S)-D[L(S)-n(S)+1]+2}{\cal O}_S \, . \nonumber
\eeqa
We have a pole at $D=4$ if $S$ is a four point subgraph. If $S$ is a two point subgraph we have poles at $D=4-2/n(S)$ and $D=4$.
As all the singularities are of this type we conclude that ${\cal A}_G$ is a meromorphic function in the strip
\beqa
{\cal D}^\sigma=\{ D\, | \, 0 < \Re\, D < 4+\e_G \}.
\eeqa

\qed

\subsection{The subtraction operator}

The subtraction operator is similar to the usual one (see \cite{reg} \cite{ren}), with the notable difference that the set of primitively divergent subgraphs are different.
We give here a brief overview of its construction. For all functions $\rho^{\nu}g(\rho)$
with $g(0)\neq 0$, denote $E(\nu)$ the smallest integer such that $E(\nu)\ge \Re \nu$.
Let
\beqa
T_{\rho}^{q}=\sum_{k=0}^q\frac{1}{k!}g^{(k)}(0), \, q\ge 0;
\quad T_{\rho}^{q}=0, \, q<0\, ,
\eeqa
be the usual Taylor operator.
We define a generalized Taylor operator of order $n$ by
\beqa
\tau_{\rho}^n[\rho^{\nu}g(\rho)]=\rho^{\nu}
T_{\rho}^{n-E(\nu)}[g(\rho)] \, .
\eeqa

To each primitively divergent subgraph we associate a subtraction operator 
$\tau_S^{-2L(S)}$ acting on an integrand like
\beqa
\tau_S^{-2L(S)}\left( \frac{e^{-\frac{HV_G}{HU_G}}}{HU_G^{D/2}}
\right)=
\Big{[}
\tau_{\rho}^{-2L(S)}
\left(\frac{e^{-\frac{HV_G}{HU_G}}}{HU_G^{D/2}}\vert_{t_S\mapsto \rho^2 t_S}
\right)
\Big{]}_{\rho=1} \, .
\eeqa

Take the example of a bubble subgraph $S$. It is primitively divergent, and 
taking into account the factorization properties we have
\beqa
\tau_{S}^{-4}\left( \frac{e^{-\frac{HV_G}{HU_G}}}{HU_G^{D/2}}
\right)=
\left(\frac{e^{-\frac{HV_{G/S}}{HU_{G/S}}}}{HU_{G/S}^{D/2}}
\frac{1}{(HU^l_S)^{D/2}}
\right)
\Big{[}
\tau_{\rho}^{-4}(\frac{1}{\rho^D})
\Big{]}_{\rho=1} \, .
\eeqa
If $D<4$ $E(D)=-3$ and if $D\ge 4$ $E(D)=-4$. Consequently 
\beqa
\tau_S^{-4}=
\begin{cases}
     {\displaystyle \hspace{2cm}0} &\text{if } D<4,
     \vspace{.3cm}\\
     {\displaystyle \left(\frac{e^{-\frac{HV_{G/S}}{HU_{G/S}}}}{HU_{G/S}^{D/2}}
    \frac{1}{(HU^l_S)^{D/2}}\right)} &\text{if } D\ge 4.
     \vspace{.3cm}\\
\end{cases}
\eeqa

As expected the operator subtracts only for $D\ge 4$, and it exactly compensates the divergence in the expression (\ref{eq:polamp}).

We then define the complete subtraction operator as
\beqa
R=1+\sum_{{\cal F}}\prod_{S\in {\cal F}}(-\tau^{-2L(S)}_S) \, ,
\eeqa
where the sum runs over all forests of primitively divergent subgraphs.

From this point onward the classical proofs of (see \cite{reg}, \cite{ren}) go through. Theorems $1$, $2$ and $3$ of \cite{ren} so that we have the theorem
\begin{theorem}
The renormalized amplitude
\beqa
{\cal A}^r_{G}=R {\cal A}_{G}\, ,
\eeqa
is an analytic function of $D$ in the strip:
\beqa
{\cal D}^\sigma=\{ D\, | \, 0 < \Re\, D < 4+\e_G \},
\eeqa
for some small positive number $\e_G$.
\end{theorem}
\section{Conclusion and perspectives - towards a non-commutative Standard Model ?}
\label{sec:conclusion}

We have presented in this paper the dimensional regularization and
dimensional renormalization for the vulcanized $\Phi^{\star 4}_4$ model. 
The factorization results we have proven are the starting point
for the implementation of a Hopf algebra structure for NCQFT \cite{progres}.

The implementation of the dimensional
renormalization program for covariant NCQFT (e.g. non-commutative Gross-Neveu or the Langmann-Szabo-Zarembo model \cite{LSZ}, see section $1$) should follow the layout presented here.

One should try to extend the techniques presented here to 
non-commutative L-S dual gauge theories. Such models have recently been proposed, 
\cite{gauge1, gauge2} but the reader should be aware that no proof of renormalizability of this models yet exists.

All the results mentioned here are obtained on a particular choice of non-commutative geometry, the Moyal space. One should try to extend this results to the NCQFT's on more involved geometries, like for example the non-commutative tori.

As mentioned in the introduction, NCQFT is a strong candidate for new
physics beyond the Standard Model. One could already study possible
phenomenological implications for Higgs physics of such non commutative renormalizable
$\Phi^{\star 4}_4$ models. The absence of Landau ghost makes this theory better behaved than its commutative counterpart. Also the Langmann-Szabo symmetry responsible for supressing the ghost could play a role similar to supersymmetry in taming UV divergencies.

\bigskip

{\bf Acknowledgment:} 
We thank Vincent Rivasseau for indicating us references \cite{reg} and \cite{ren}
and for fruitful discussions during the various stages of the preparation of this work.

\end{document}